\documentstyle[11pt,newpasp,twoside,epsf]{article}
\newcommand{\jj}[2]{\mbox{$J = #1\rightarrow#2$}}
\newcommand{\kms}{\mbox{km s$^{-1}$}}
\newcommand\cmv{\mbox{cm$^{-3}$}}

\newcommand\submm{submillimeter}

\newcommand{\lsun}{\mbox{L$_\odot$}}
\newcommand{\lsub}{\mbox{L$_{submm}$}}
\newcommand{\msun}{\mbox{M$_\odot$}}
\newcommand{\tk}{\mbox{$T_K$}}
\newcommand{\td}{\mbox{$T_D$}}
\newcommand{\Tdr}{\mbox{$T_D(r)$}}
\newcommand{\tbol}{\mbox{$T_{bol}$}} 

\newcommand{\mean}[1]{\mbox{$\langle#1\rangle$}} 

\newcommand{\water}{H$_2$O}
\newcommand{\methanol}{CH$_3$OH}

\begin{document}

\title{Early Phases and Initial Conditions for Massive Star
         Formation}
\author{Neal J. Evans II, Yancy L. Shirley, Kaisa E. Mueller, \& Claudia Knez}
\affil{Department of Astronomy, The University of Texas at Austin,
       Austin, Texas 78712--1083}

\begin{abstract}
Our knowledge of the initial conditions and early stages of high mass star
formation is very limited. We will review recent surveys of regions
in the early stages of massive star formation using molecular tracers of
high density and dust continuum emission and consider the status of
evolutionary schemes. Comparison to the situation for low mass, relatively
isolated star formation will be used to illustrate the outstanding issues
in massive star formation. The problem of initial conditions is
particularly acute because there is a lack of observational evidence
for regions capable of forming massive stars {\it before} star formation
actually begins. By analogy with the Pre-Protostellar Cores (PPCs) studied
for low-mass star formation, one might call such regions
Pre-Proto-cluster Cores (PPclCs). We will conclude with some speculation
about what such cores might look like and possibilities for their
detection.
\end{abstract}

\section{Introduction}

Understanding the formation of massive stars is crucial to an
understanding
of the formation of stars in general and of galaxies. 
Many, perhaps most, stars form in
clustered environments where massive stars are present (Lada 1992, 
Williams \& McKee 1997).
Only massive star formation is readily studied at the distances of other
galaxies.
In particular, the discovery that star formation at substantial redshifts 
is often heavily obscured by dust (e.g., Frayer et al. 2000; 
Chapman et al. 2001) indicates that study of the history of
star formation in galaxies, hence the major part of galaxy formation, can
be related to what we learn about massive star formation in our own and
nearby galaxies.

Here we consider a number of issues surrounding the formation of massive
stars, beginning with a comparison to what we know about the formation
of low mass stars, proceeding through some systematic studies of 
early stages of massive star formation, and ending with some speculations
about the initial conditions for massive star formation. 

\section{An Evolutionary Paradigm? }

In the case of low mass stars, a well-established evolutionary
paradigm prevails, and a standard model of star formation (Shu et al.
1987)
provides a detailed target for observational tests. Various alternative
models have also been proposed and observers are beginning to compare
these
models to observations. The evolutionary paradigm is based on the 
class system (Lada \& Wilking 1984, Lada 1987),
originally proceeding from Class I (deeply embedded) to III (revealed
star, weak-line T Tauri star) via Class II (roughly the same 
as classical T Tauri stars). 

The recent advances in submillimeter
capability have revealed a self-luminous phase that is still more embedded
than Class I, and Andr\'e, Ward-Thompson, \& Barsony (1993)
called this phase Class 0. 
The initial conditions for low mass star formation appear to be
amenable to study in dense regions without any evidence for internal
luminosity. Variously called Pre-Protostellar Cores (PPCs), 
pre-stellar cores, or Class $-1$ sources, these objects first
became obvious as peaks of \submm\ emission. While they roughly
coincide with molecular emission regions, the molecular line emission
did not directly reveal the high densities, apparently because
of substantial molecular depletion (e.g., Caselli et al. 2001), 
caused by the high densities and
low temperatures in the cores of these regions. The temperatures
are low enough ($\td \sim 7$ K) that even the \submm\ emission is 
decreased toward the centers (Evans et al. 2001), but the dust continuum
emission is much more revealing than the molecular line emission that
has usually been used to trace dense regions. 

The extended class system relies on a mixture of slopes (Lada \& Wilking
1984, Lada 1987), peaks, and ratios (Andr\'e et al. 1993)  
in the spectral energy
distribution (SED) of dust emission. An alternative that unifies these
methods and provides a continuous variable is the bolometric temperature
(\tbol ) introduced by Myers \& Ladd (1993).  
Chen et al. (1995) showed that the
class boundaries corresponded well to certain values of \tbol.
However, \tbol, being defined by the spectrally averaged mean frequency,
can be heavily affected by emission at relatively short wavelengths, where
geometrical effects are conflated with evolutionary effects.
For the early stages, there are two important boundaries: Class $-1$ to
Class 0; and Class 0 to Class I. There are usually insufficient
measurements to define \tbol\ for Class $-1$ sources; the test for Class $-1$
status is that there is no evidence for internal luminosity. 
For the Class 0/I boundary,
one can use either $\tbol = 70$K (Chen et al. 1995) 
or the ratio of emission beyond 350 \micron\
to total luminosity, $\lsub / L$ (Andr\`e et al. 1993).

To what extent can we carry the observational paradigm and theoretical
models for low-mass star formation over to the case of massive star
formation?
Working with a modest sample size (14), van der Tak et al (2000) found
rather poor correlations between different potential tracers of age. 
It is not very surprising that the SED may not correlate with evolution
in the same way as for low mass stars; for massive stars forming in 
clusters, the SED will be dominated by the most luminous object that is
still embedded and the evolution of the various objects is unlikely to 
be highly synchronized. The Class II phase in particular may have little
meaning:
for low mass stars, this is the phase when the SED is dominated by a disk;
the presence of disks around massive stars is less firmly established.
Finally, the evolution of massive stars is so much faster that we cannot
expect the gentle unfolding of the SED as the dust moves from envelope to
disk to star in the low mass case. Instead, massive stars reach the
main sequence, form HII regions, blow ionized winds, etc. before much
of the material in the envelope has fallen in.

What other evolutionary tracers might prove useful? Clearly, the
emergence of an HII region, first as an ultra-compact HII region,
marks an important evolutionary boundary. If we are looking for the
earliest stages, we want to study sources before UC HII regions form.
There are two general ideas for evolutionary tracers at earlier times:
masers and chemical signatures. While OH masers appear to be associated
with
HII regions, \water\ and possibly \methanol\ (e.g., Minier, Conway, \&
Booth
2001) masers appear to be
associated with early stages, when molecular outflows produce shocks.
Hot cores (warm, dense, chemically-rich regions) also appear to precede
UC HII regions (see van der Tak 2002 for a discussion of this approach).  
The hot cores could in principal be heated internally
(and hence analogous to Class 0 sources) or externally (hence analogous
to Class $-1$ sources).

There have been many detailed studies of individual regions of massive
star formation, but systematic studies of large samples of regions using
uniform analytical methods are just becoming available (e.g., Plume et al.
1992, 1997, Bronfman et al. 1996, Sridharan et al. 2001). In the next section,
we report on recent work on the sample identified by Plume et al. (1992).

\section{Studies of Regions Associated with Water Masers}

Water masers provide a large sample of objects with accurate
positions for further study (e.g., Cesaroni et al. 1988). 
Many masers are {\it near} UC HII regions, but their positions do not 
coincide, consistent with the masers revealing an earlier stage. 
In addition, the presence of a \water\
maser ensures the existence of some very dense gas. The models of
these masers also require shocks, suggesting that some star formation
and associated outflow has begun. 
Consequently, we have
pursued a series of studies of the gas and dust in the directions of
water masers, beginning with a survey for emission in the \jj76
transition of CS (Plume, Jaffe, \& Evans 1992). 
A large fraction of the maser sources
were detected in the CS \jj76 survey, leading to a follow-up study
using CS \jj21, \jj32, \jj54, and in a few cases, $J = 10\rightarrow 9$
transitions to constrain the density via LVG calculations of the molecular
excitation (Plume et al. 1997). For many sources, C$^{34}$S data provided
a check on optical depth effects in CS. That study found high densities
in many sources, with a mean in the log of density, $n(\cmv)$, of
$\mean{{\rm log} n} = 5.9$, with the average taken over 71 sources.

Plume et al. (1997) also mapped a few sources crudely (cross-scans) 
to estimate sizes
and masses. To obtain better data on sizes, masses of dense gas, density
distributions, etc., we have made fully sampled maps of a large fraction
of the
Plume et al. sample. We mapped 63 regions in CS \jj54 (Shirley et al.
2002) and 24 regions in CS \jj76 (Knez et al. 2002). 
In addition, we have mapped the 350 \micron\ continuum emission from dust 
toward 51 of these regions (Mueller et al. 2002). 
Details of the observations and
analysis can be found in those papers, and a more complete study of
one of the objects can be found in Lee et al. (2002). Here we focus
on the overall results, summarizing and
comparing the results from the different tracers, while accounting for 
differences among the samples in the different studies.

For all the samples, the FWHM size of each core was determined by 
deconvolving the telescope main beam FWHM from the observed FWHM of the
core. 
In most cases, the deconvolved angular sizes determined in this way were 
slightly larger than the beam size, characteristic of a power-law emission
distribution (Terebey et al. 1993). 
Also, the CS \jj76 sizes were smaller than the CS \jj54
sizes, as expected for a centrally peaked density distribution because
the critical density for the \jj76\ line is 3 times greater than that for the
\jj54 line (cf. Evans 1999).
For power law distributions, the size defined in this way should 
not be considered a physical boundary. Instead, we use it, along with
the distance, to calculate a fiducial radius for convenient comparisons.

Similarly, the mass in a power law distribution can only be specified for
a given radius; when comparing masses from various techniques, it is
important to compare for the same fiducial radius. Because the CS \jj54\
maps have the largest extents, thus tracing the core farther down the density
distribution, we compare all masses calculated to be
inside the radius of the CS emission: $r_{CS} = (D/2) \theta_s$, where
$\theta_s$ is the FWHM angular extent of the map of CS \jj54\ emission and
$D$ is the distance.

\subsection{Models}
 
\begin{figure}[h!]
\plotfiddle{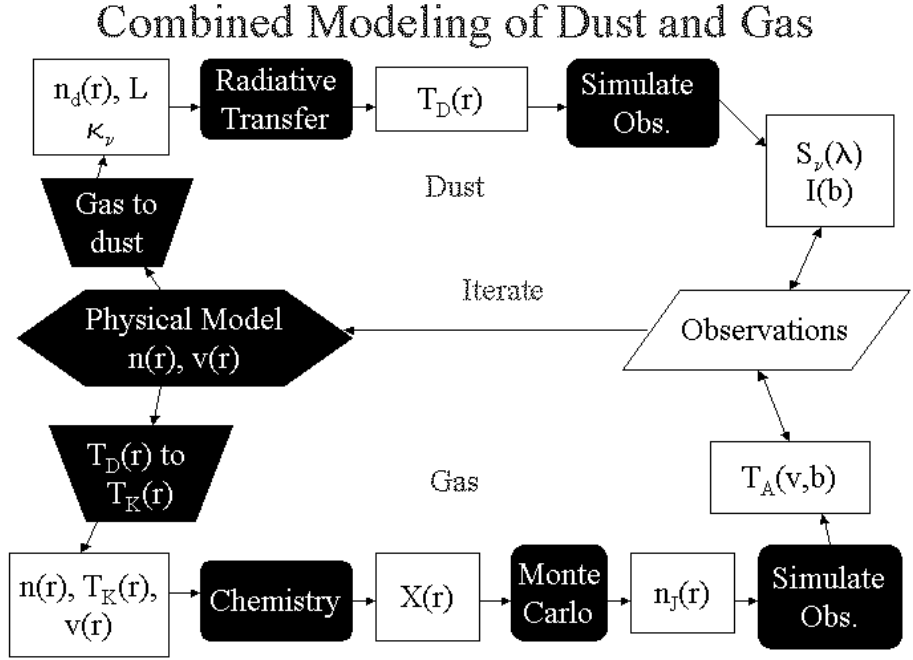}{3.8in}{-90}{50}{50}{-200}{300}
\caption{The modeling scheme is shown, with the modeling of dust emission
following the top loop and the modeling of emission from gas following the
bottom loop. Not all steps are fully implemented yet.
}
\end{figure}

The modeling follows the scheme represented by Figure 1, though at present
some of the processes are incompletely implemented. 
The modeling of the dust
emission is described by Mueller et al. (2002). A density
power law ($n(r) = n_0 (r/r_0)^{-p}$) is assumed, a radiative transport
code
is used to compute a self-consistent \Tdr, and the resulting \Tdr\ and
$n(r)$ are used to compute the dust emission over the full 
SED. For comparison with maps of
\submm\ continuum emission, the emission is convolved with the measured
beam response and chopping is simulated. Comparison of the predicted 
intensity profile to that obtained in the observations allows one to 
choose the best fitting power law index for the density. Comparison to
the observed integral of the SED constrains the internal luminosity of the 
forming stars, while comparison to the emission 
at \submm\ wavelengths constrains
$n_0$ and hence the mass. The mass also depends on the assumed opacity
at 350 \micron\ ($\kappa_{\nu}(350)$), and the overall shape of the SED 
depends on the full behavior of the dust opacity as a function of
frequency ($\kappa_{\nu}(\nu)$). 
We assume the dust opacities given in column  5
of Ossenkopf and Henning (1994), hereafter OH5, but divide by 100 so that
are given per gram of total mass (gas plus dust). These opacities have
fit the SEDs from both low-mass (Evans et al. 2001, Shirley et al. 2001) and
high-mass (van der Tak et al. 2000) star formation regions.
They are appropriate for grains that have coagulated at high density and
accreted thin ice mantles. Different choices for opacities make little
difference in the optimum $p$, but they do affect the mass because only
the product $M \kappa_{\nu}(350)$ enters the equation for the dust
emission.

\begin{figure}[h!]
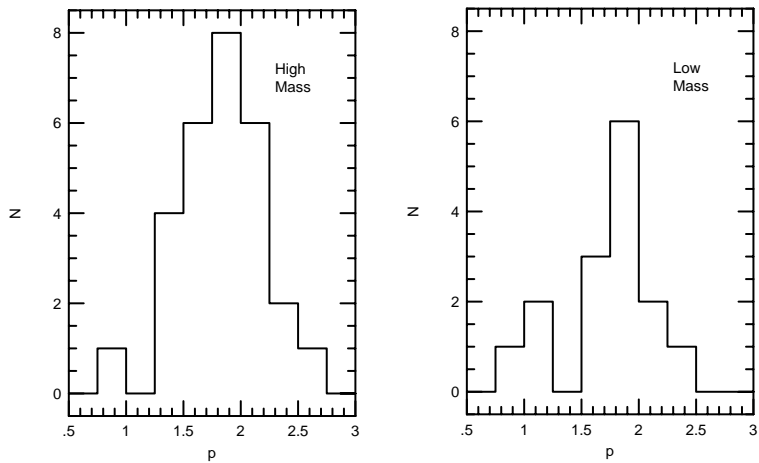

\plotfiddle{phist2.eps}{145pt}{0}{25}{25}{-150}{-25}
\plotfiddle{phist.eps}{0pt}{0}{25}{25}{0}{0}
\caption{Histograms of $p$, best-fit density power law exponents for
high-mass star forming regions on the left (Mueller et al. 2002) and
for low-mass star forming regions on the right (Shirley et al. 2001,
Young et al. 2001). All are based on modeling the radial distribution
of the submillimeter continuum emission from dust.
}
\end{figure}

For the 28 sources with models, Mueller et al. (2002) find
a mean density index of $\mean{p} = 1.72\pm 0.37$, with a distribution
shown in Figure 2 (left panel). 
This value is considerably larger than that found by van der Tak et al.
(2000), who modeled a subset of these sources. For that earlier work,
the beam was approximated as a Gaussian, while the observed beam,
including
azimuthally averaged sidelobes, was used in the models by Mueller et al.
(2002). For sharply peaked intensity distributions, the details of the
beam profile make a substantial difference; source-by-source comparison
indicates that Mueller et al. find values of $p$ larger than those found
by van der Tak et al. by about 0.4 on average.
The \mean{p}\ found by Mueller et al. (2002) is consistent within
the uncertainties with that found by Beuther et al. (2001) ($\mean{p} = 
1.6\pm 0.5$) for a differently-selected sample of regions forming massive
stars. These power-law indices are also similar in the mean to those found for
isolated cores forming low mass stars (Shirley et al. 2000, 2001; Young et al.
2001; Motte \& Andr\'e 2001).  In Figure 2, right panel, we show the
histogram for the combined sample of low-mass cores modeled by Shirley et al.
(2001) and Young et al. (2001), for which $\mean{p} = 1.63\pm 0.40$. 
While the means agree, the distributions
may be different, though larger samples are still needed. The low-mass
regions primarily have $p \sim 2$; the smaller peak
between 1 and 1.5 is composed of sources that are quite aspherical.
The distribution of $p$ for regions forming massive stars appears to
have a larger intrinsic spread.

The detailed modeling of the CS emission (Knez et al. 2002)
is at an early stage, with only
one source (M8E) modeled, and some of the processes in the bottom loop
of Figure 1 are incompletely implemented.
In particular, we use the \Tdr\ from the dust modeling and 
currently assume $\tk = \td$ (a good assumption at the densities probed by CS
emission). The turbulent velocity dispersion is assumed to be constant,
and no systematic motions (e.g., collapse) are assumed.
Instead of running a chemical model, we assume a constant CS abundance,
which is adjusted to match the optically
thin emission from C$^{34}$S \jj54. Various values of $p$ were used to
allow a determination independent from that of the dust emission. For M8E, 
the CS prefers a slightly lower value of $p$ than the dust: 1.6 rather 
than 1.75. Given the uncertainties and the assumptions, these are in quite
good agreement.

\subsection{Masses}

\begin{figure}[ht!]
\plotfiddle{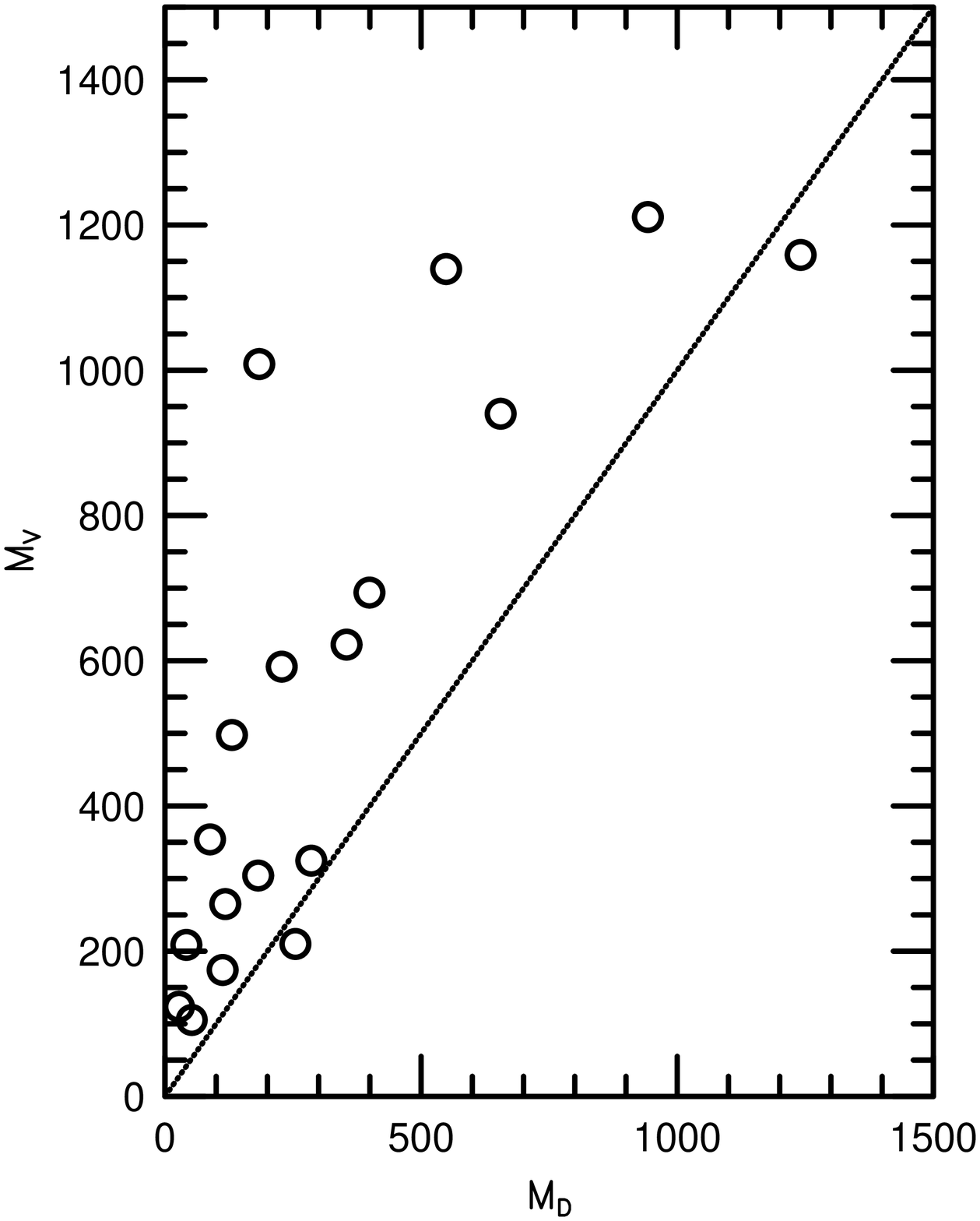}{150pt}{0}{30}{30}{-180}{-50}
\plotfiddle{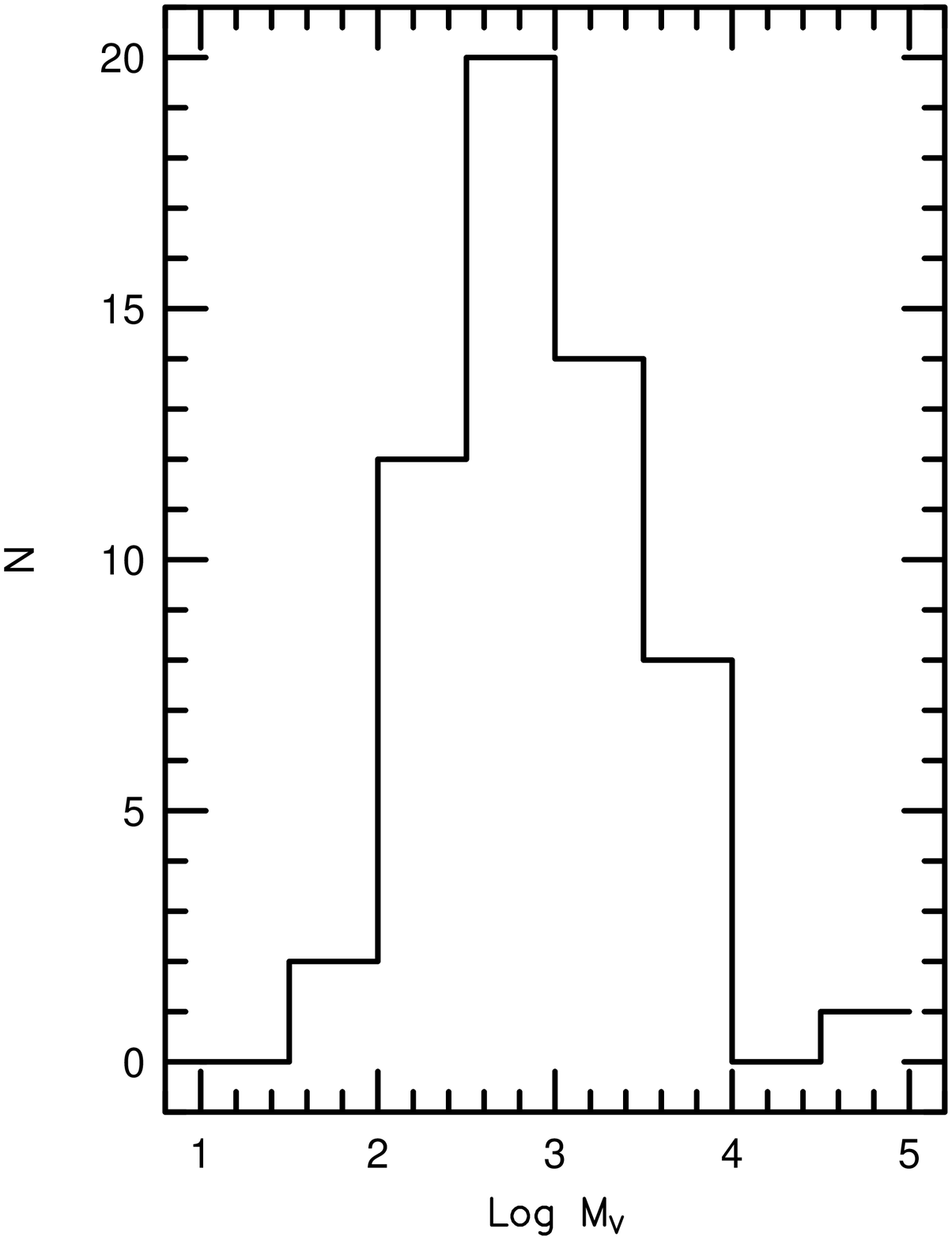}{0pt}{0}{30}{30}{10}{-20}
\caption{The corrected CS virial mass (density power law and linewidth
broadening corrected) compared to the total (gas plus dust) mass based on 
models of the dust emission for 18 sources (left) and the
corrected CS virial mass histogram for the entire sample of 57 sources
(right).
}
\end{figure}

Mueller et al. (2002) find a mean mass of gas and dust based on the dust
emission within the FWHM of the 350 \micron\ emission, $\langle M_D
\rangle =  209 \msun$. 
Because the \submm\ emission
has the smallest beam, and because the \submm\ sample does not include
the more massive cores in the full sample, this mass is much less than
the mean virial mass, $\langle M_V \rangle =  4550 \msun$, 
determined from the CS emission (Shirley et al. 2002). In addition, that
estimate of virial mass assumed constant density and used the FWHM velocity
width of the CS \jj54 line, which is broadened by optical depth effects.
We can correct the virial mass $M_V$ for
the density power law index, according to (Bertoldi \& McKee 1992),

\begin{displaymath}
M_V = \frac{5(1 - \frac{2p}{5}) r_{CS} \Delta v^2}{8 {\rm ln} 2
G(1 - \frac{p}{3})}
\end{displaymath} 
where the factor 8ln2 comes from converting from mean square velocity to
FWHM linewidth ($\Delta v$), and 
$r_{CS}$ is the radius defined above for the CS \jj54 emission. 
For this calculation, we use the FWHM of the C$^{34}$S line,
which corrects
for saturation broadening of the CS lines.  Preliminary comparisons of
$^{13}$CS and C$^{34}$S spectra indicate that C$^{34}$S \jj54 linewidths 
are not broadened due to optical depth effects.

For comparison, we calculate the total mass based on the model of the 
dust emission, but integrated
out to the same radius ($r_{CS}$) as used for the virial mass:
\begin{displaymath}
M_D = \frac{4 \pi \mu m_H n_0}{r_0^{-p}} \int_0^{r_{CS}} r^{2-p}\, {\mathrm d}r
= \frac{4 \pi \mu m_H n_0 r_{CS}^{3-p}}{(3-p) r_0^{-p}} 
\end{displaymath}
where $\mu m_H$ is the mean molecular mass and $n_0$ is the density at
$r_0$.
Because the value of $n_0$ is fixed by matching the observed flux density
at 350 \micron, $n_0 \propto 1/\kappa_{\nu}(350)$.
The ratio of the virial mass to the total (gas and dust) mass determined
from the dust modeling then depends on measured and assumed quantities
as follows:
\begin{displaymath}
M_V/M_D \propto 
\Delta v^2  r_{CS}^{p-2}/n_0 \propto
\Delta v^2 \kappa_{\nu}(350) r_{CS}^{p-2}
\end{displaymath}
For the 18 sources with all the required data and models,
$\mean{M_{V}/M_D} = 2.4 \pm 1.4$ (see Figure 3). 
Because the opacities from different dust models vary by up to a factor of 10
(e.g., Ossenkopf \& Henning 1994), this discrepancy is not very large.
The agreement of the two very different mass estimates at this level 
suggests that both the assumptions
underlying the virial mass estimate and the OH5 opacities at 350 \micron\
are quite good. A factor of about 2 decrease in $\kappa_{\nu}(350)$ from the 
OH5 value would bring them into agreement on average. 

Based on the good agreement of the masses on average, we calculated the
virial masses for the rest of the sample, assuming the mean value of $p$
and correcting for linewidth broadening in the CS \jj54\ line 
statistically (Shirley et al. 2002). 
The resulting mass distribution has a mean value of $2030 \pm 620$ \msun. 
Recall that this number refers to the mass inside $r_{CS}$; if the core
extends to twice this radius with the same power law in density, the 
estimate of $M_V$ grows by a factor of 2, while the mass from dust emission
grows by $2^{3-p}$ or a factor of 2.4 for the mean $p$, leading to 
estimates of 4000 to 4900 \msun, for the mean mass.
While the mass histogram appears peaked, selection effects against small
sources are present at the mean distance of sources in this sample:
$5.5\pm 3.7$ kpc. For sources above 1000 \msun, the distribution could be 
fitted by a power law in mass with an exponent of 1.8 to 2.0 (Shirley et al.
2002).

\subsection {Evolutionary Indicators}

Evolutionary indicators ($T_{bol}$, $L_{smm}/L_{bol}$, and $p$)
for the water maser sample and for low mass protostars observed with SCUBA 
at 850 and 450 $\micron$ (Young et al. 2001, Shirley et al. 2001, 
Evans et al. 2001, Shirley et al. 2000) are plotted in Figure 4.  
Based on the evolutionary scheme for low mass star formation, 
$T_{bol}$ would be expected to 
increase  while $L_{smm}/L_{bol}$ would decrease
as the envelope dissipates.  Indeed, for low mass protostars, 
these trends seem to hold for a small sample  of sources in
 the most embedded phases of evolution (PPC through Class I).  
The water maser sources weakly follow a similar trend; however, the
water maser sample does not cover as wide a range in $T_{bol}$
and $L_{smm}/L_{bol}$ as the low mass core sample.  
Preliminary analysis of dust models for both high and low mass cores
do not indicate a strong correlation of the density power law index,
$p$, with $T_{bol}$. Both samples are biased towards more embedded 
regions and therefore sample a restricted range of $T_{bol}$ and 
$L_{smm}/L_{bol}$.  Each indicator has a different dependence of geometrical
effects, such as outflow cavities, clumping, etc.
$T_{bol}$ and $L_{smm}/L_{bol}$ are determined from the SED, while
the density power law index, $p$, probes the density distribution
in a different way.
While larger samples covering a wider range of the evolutionary parameters
are needed, it seems that the traditional evolutionary indicators for
low mass star formation may have some utility for massive stars, at least
at the early (Class 0 and I) stages.

\begin{figure}
\plotfiddle{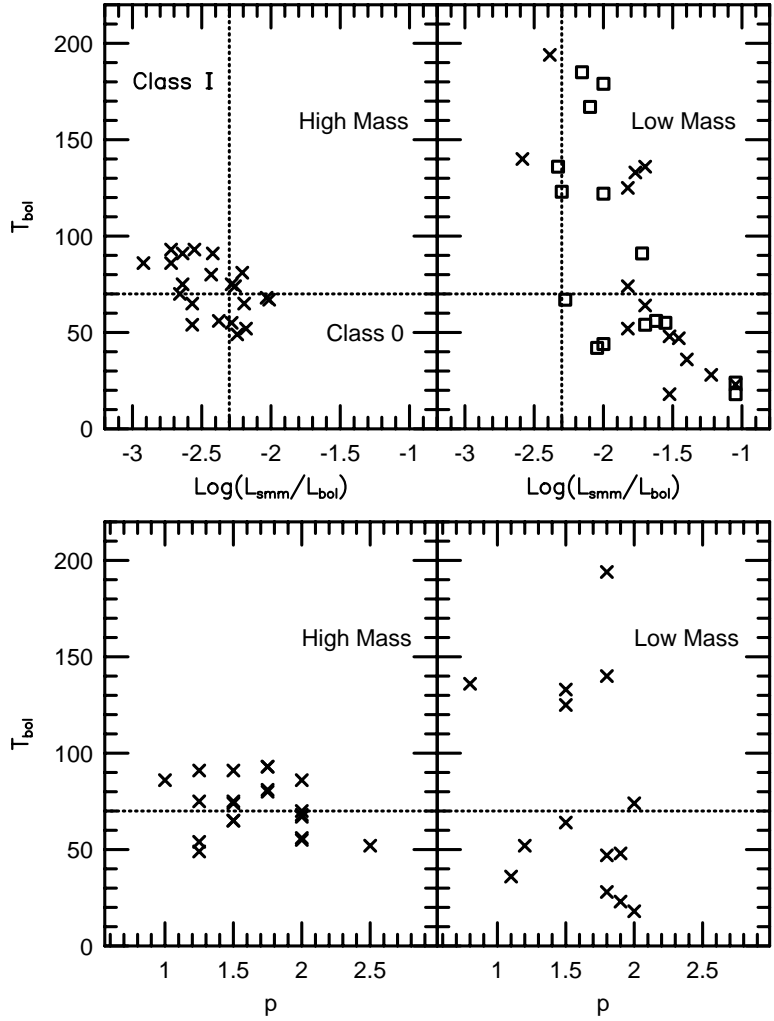}{358pt}{0}{58}{58}{-180}{-45}
\caption{Potential evolutionary indicators for low and high mass star forming
regions. The crosses indicate sources with models, while the open boxes
in the upper right diagram represent sources that have not been modeled.
The dashed lines indicate the boundary between Class 0 and Class I 
protostars in the low mass evolutionary scheme (Class 0 sources
should have $\tbol < 70$ K and $L_{submm}/L_{bol} > 0.005$).
Note that the two tracers (\tbol\ and $L_{submm}/L_{bol}$) disagree for
a substantial number of the low mass cores, suggesting that the dividing
lines may need revision.
}
\end{figure}

\subsection{Summary}

The results so far from the study of water maser sources indicate that
conditions in regions forming massive stars are quite different from those
in regions forming low mass stars. The total mass of dense ($n \sim 10^6$
\cmv) gas is much greater, the turbulence is much higher for a given size,
and, once stars have formed, the temperature is much higher.
For the subset of sources with adequate data, the luminosity to mass
ratio is $192 \pm 145$ $\lsun/\msun$ for the mass of {\it dense} gas determined
from the virial mass within $r_{CS}$. 
For the masses based on the dust emission,
the same ratio is $334 \pm 231$ $\lsun/\msun$. 
The latter number may be more relevant for comparison to observations of
distant galaxies where long wavelength dust emission is used to trace mass.
These numbers are for 18 sources that have models for the dust emission,
and are still preliminary until more sources can be added.

These values for $L/M$ are much higher than that inferred from CO studies,
which provide masses for the whole molecular cloud. The average value
from CO studies is $\mean{L/M} = 4$ \lsun/\msun\ (Mooney \& Solomon 1988)
and the distribution spreads over a factor of $10^2$ (Evans 1991).
The dense gas provides a much
better guide to the star formation rate and efficiency, consistent with
the conclusions of Gao \& Solomon (2000), 
who showed that HCN correlates much better than CO 
with star formation indicators in a large sample of galaxies.
The $L/M$ for dense gas is comparable to the values inferred for starburst
galaxies, based on CO emission: $L/M \sim 100$ (Kennicutt 1998, 
Sanders et al. 1991).
Starburst galaxies appear to form stars as if a large fraction of their
molecular gas content behaves like the dense cores associated with water
masers.

\section{The Initial Conditions}

So far, all the regions studied already have ongoing star formation.
What initial conditions lead to the conditions in the observed regions?
By analogy with the Pre-Protostellar Cores (PPCs) for low mass star
formation,
the initial conditions for massive, clustered star formation might be
referred to as 
Pre-Proto-cluster Cores (PPclCs, an acronym chosen deliberately to avoid
its future use).
Based on the mean properties of the \water\ maser sample, we would expect
the following properties. They would be dense ($n \sim 10^6$ \cmv) and 
massive ($M \sim 5000$ \msun).
The linewidth is a question: if the turbulence is stirred up by star
formation,
it could be less before star formation begins; however, the virial
linewidth
for a core of the mean mass would be 7.5 \kms.
Before an internal luminosity source develops, the PPclC should be very
cold
on the inside because it is heated only externally by the interstellar 
radiation field (ISRF). Studies of low-mass PPCs indicate \td\ dropping
to 7--8 K at the center (Evans et al. 2001, Zucconi et al. 2001).

So far, we have no clear examples of such objects! There are less massive
candidates, but nothing we have found in the literature 
approaches these properties. 
Why do we lack examples? There are several possible explanations:
\begin{enumerate}
\item PPclCs will be very rare.
\item Appropriate searches have not been made.
\item PPclCs never exist.
\end{enumerate}

The first explanation arises because massive star formation is rare
in general and because PPclCs should evolve rapidly to form stars.
The second explanation can certainly be part of the answer. The low mass
PPCs were found only recently, using \submm\ continuum observations of
starless cores originally identified by visual obscuration. While
molecular line emission identified the general region, the high densities
in the PPCs became apparent only with \submm\ dust emission studies.
The molecular tracers fail to trace the densest regions of PPCs,
apparently because of  a combination of opacity and heavy
depletion onto dust (e.g., Kramer et al. 1999, Caselli et al. 2001). 
Because no obvious
signposts like maser emission mark the PPclCs, unbiased \submm\ surveys
of large regions of distant molecular clouds are needed to find them.
While such surveys are becoming possible and should be attempted, it is
only fair to note the third explanation. Regions forming clusters may
never go through a PPclC phase if lower-mass stars form while the
full mass is being assembled. In that case, one might see various 
signposts of lower mass star formation (outflows, infrared emission)
associated with a {\it cluster}
of cores that are merging to form the truly massive cores 
observed at later stages. 

Some possible candidates for PPclCs may be found from large scale
surveys. For example, Lada et al. (1991) surveyed 3.6 square degrees
of the L1630 cloud in CS \jj21\ emission, finding numerous dense
cores. The large cores all had ongoing star formation; follow-up 
studies (Lada et al. 1997) showed that the starless cores were all
of modest mass ($M \sim 10 \msun$). The most promising candidates
at present are the mid-infrared-dark clouds found by the MSX mission
that lack IRAS emission
(Egan et al. 1998). Follow-up studies using H$_2$CO (Carey et al.
1998a) found massive ($M \sim 10^3$ \msun), cold ($\td \sim 10$ K)
cores with $n \sim 10^6$ \cmv. Maps of the submillimeter dust emission
(Carey et al. 1998b)
showed a string of cores, some of which harbor molecular outflows.
The outflows indicate that some star formation may have begun, but these
objects at least seem to be at a very early evolutionary stage.

\subsection{A Model for the PPclC}

\begin{figure}
\plotfiddle{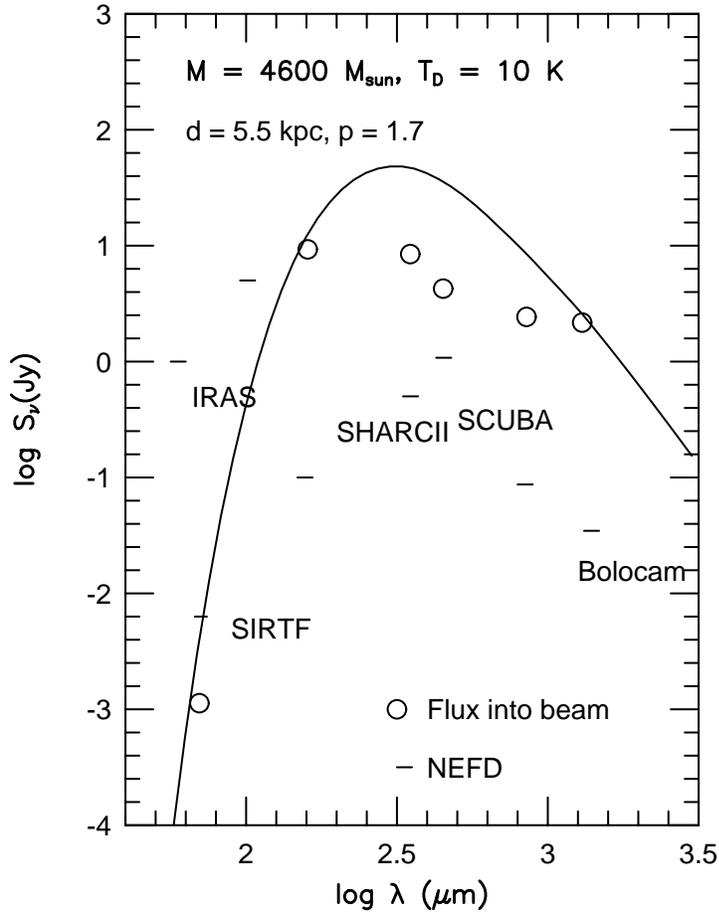}{5.0in}{0}{50}{50}{-150}{0}
\caption{The SED of a putative Pre-Proto-cluster Core, based on typical
properties of the cores found around water masers, but assuming a constant
dust temperature of 10 K. The emission from the overall core is shown by
the solid line, while the flux density contained with a single beam of
various instruments is shown as open circles. The horizontal bars show the
NEFD of these instruments.
}
\end{figure}

To provide a target for possible searches, we have constructed a model
of a PPclC based on the mean properties of the \water\ maser sample:
$M = 4600 \msun$, $p = 1.72$, and $d = 5.5$ kpc.
The outer radius of the model was taken to be 0.76 pc, twice the
mean radius from the CS maps. At the mean distance of 5.5 kpc, this radius
translates to $\theta_s =  56\arcsec$. 
The dust temperature was assumed constant at 10 K; while models of
PPCs indicate lower \td\ in the center, PPclCs may exist in regions of
higher radiation fields. OH5 opacities were assumed.
Observations of the model with various past, current, and future 
instruments were simulated. The results are shown in Figure 5.
The solid line is the total emission, while the circles show the
emission into the beam of the instrument. The horizontal bars show the
limits for 1 sec of integration. It is not surprising that
IRAS did not detect these objects; SIRTF could detect them at 160 \micron,
but that channel will be saturated for most regions of molecular gas just 
from the extended cloud emission. The best regime in which to search 
for such objects is the millimeter to submillimeter, and systems with
fairly
large beams and fields of view are ideal for the large-scale blind surveys
that will be needed. A good example of such an instrument is Bolocam, 
operating on the CSO (Glenn et al. 1998).

\section{Summary}

The formation of massive stars occurs in parts of molecular clouds that
are much denser and more turbulent than the bulk of the molecular gas.
Compared to the regions forming low mass stars, the birthplaces of massive
stars have high densities over much larger regions and their linewidths
are much greater than those inferred from the linewidth-size relation.
A clear evolutionary sequence is difficult to establish, but regions 
surrounding water masers should be younger than those around ultra-compact
HII regions. Studies based on \water\ masers find a distribution of
masses of dense ($n \sim 10^6$) gas that extends to quite large masses.
While affected by selection biases, a mean value of 2000--4000 \msun\
is found, enough to form large clusters of stars. Modeling the 
radial profiles of \submm\ emission from dust indicates that power laws in
density ($n(r) \sim r^{-p}$) fit the data with a mean $p = 1.72$, rather
similar to that found for molecular cores around ultra-compact HII regions
(Beuther et al. 2001) and to low mass star-forming cores (Shirley et al.
2001; Young et al. 2001).

\acknowledgments
This work has been supported by NASA grants NAG5-7203 and NAG5-10488,
by NSF grant AST-9988230, and by the State of Texas.


\begin{references}

\reference
\reference Andr\'e, P., Ward-Thompson, D., \& Barsony, M. 1993, \apj, 406,
122
\reference Bertoldi, F., \& McKee, C. 1992, \apj, 395, 140
\reference Beuther, H., Schilke, P., Menten, K. M., Motte, F., 
    Sridharan, T. K., \& Wyrowski, R. 2001, \apj, in press
(astro-ph/0110370)
\reference Bronfman, L. Nyman, L.-A., May, J. 1996, \aaps, 115, 81
\reference Carey, S.~J., Clark, F.~O., Egan, M.~P., Price, S.~D., 
   Shipman, R.~F., \& Kuchar, T.~A. 1998a, \apj, 508, 721 
\reference Carey, S.~J., Redman, R.~O., Feldman, P.~A., Egan, M.~P., \& 
    Shipman, R.~F.\ 1998b, American Astronomical Society Meeting, 193, 6525 
\reference Caselli, P., Walmsley, C. M., Zucconi, A., Tafalla, M., Dore,
L., 
    \& Myers, P. C. 2001, \apj , in press (astro-ph/109023)
\reference Cesaroni, R., Palagi, R., Felli, M., Catarzi, M., Comoretto,
G.,
    DiFranco, S., Giovanardi, C., \& Palla, F. 1988, \aaps, 76, 445
\reference Chapman, S. C., Richards, E. A., Lewis, G. F., Wilson, G., \&
    Barger, A. J. 2001, \apj, 548, L147
\reference Chen, H., Myers, P. C., Ladd, E. F., \& Wood, D. O. S. 1995,
   \apj, 445, 377
\reference Egan M.~P., Shipman, R.~F., Price, S.~D., Carey, S.~J., Clark, 
     F.~O., \& Cohen, M. 1998, \apjl, 494, L199
\reference Evans, N. J., II 1991, in Frontiers of Stellar Evolution, ed.
  D. L. Lambert, (San Francisco:Astron. Soc. Pacific), 45 
\reference Evans, N. J., II 1999, \araa, 37, 311
\reference Evans, N. J., II, Rawlings, J. M. C., Shirley, Y. L., \&
   Mundy, L. G. 2001, \apj, 557, 193
\reference Frayer, D. T., Smail, I., Ivison, R. J., \& Scoville, N. Z.
         2000, \aj, 120, 1668
\reference Gao, Y. \& Solomon, P. 2000, American Astronomical Society 
    Meeting, 197, 9603
\reference Glenn, J. et al. 1998, SPIE, 3357, 326
\reference Kennicutt, R. C., Jr. 1998, \araa, 36, 189
\reference Knez, C., Shirley, Y. L., Evans, N. J., II, \& Mueller, K. E., 
    2002, these proceedings
\reference Kramer, C., Alves, J., Lada, C. J., Lada, E. A., Sievers, A., 
    Ungerechts, H., \& Walmsley, C. M. 1999, \aap, 342, 257
\reference Lada, C. J. 1987, in Star Forming Regions, IAU Symp. 115, ed.
    M. Peimbert \& J. Jugaku (Dordrecht: Reidel), 1
\reference Lada, C. J., \& Wilking, B. A. 1984, \apj, 287, 610
\reference Lada, E. A. 1992, \apjl, 393, L25
\reference Lada, E. A., Bally, J., \& Stark, A. A. 1991, \apj, 368, 432
\reference Lada, E. A., Evans, N. J., II, \& Falgarone, E. 1997, \apj, 488, 286
\reference Lee, J.-E., Young, C. H., Shirley, Y. L., Mueller, K. E., \& 
    Evans, N. J., II 2002, these proceedings
\reference Minier, V., Conway, J. E., \& Booth, R. S. 2001 \aap, 369, 278
\reference Mooney, T. J., Solomon, P. M. 1988, \apjl, 334, L51
\reference Motte, F., \& Andr\'e, P. 2001, \aap, 365, 440
\reference Mueller, K. E., Shirley, Y. L., Evans, N. J., II, \& 
   Jacobson, H. R.  2002, these proceedings
\reference Myers, P. C., \& Ladd, E. F. 1993, \apjl, 413, L47
\reference Ossenkopf, V., \& Henning, Th. 1994, \aap, 291, 943
\reference Plume R., Jaffe, D. T., Evans, N. J., II  1992, ApJS, 78, 505
\reference Plume R., Jaffe, D. T., Evans, N. J., II  Martin-Pintado, J., 
           \& Gom\'ez-Gonz\'alez, J. 1997, ApJ, 476, 730
\reference Sanders, D. B., Scoville, N. Z., \& Soifer, B. T. 1991, \apj, 
	   370, 158
\reference Shirley, Y. L., Evans, N. J., II, Mueller, K. E., Knez, C.,
   \& Jaffe, D. T. 2002, these proceedings
\reference Shirley, Y. L., Evans, N. J., II, \& Rawlings J. M. C.,
    \& Gregersen, E. M. 2000, \apjs, 131, 249
\reference Shirley, Y. L., Evans, N. J., II, \& Rawlings J. M. C. 2001, 
   in prep.
\reference Shu, F. H., Adams, F. C., \& Lizano, S. 1987, \araa, 25, 23
\reference Sridharan, T. K., Beuther, H., Schilke, P., Menten, K. M.,
      \& Wyrowski, F. 2001, \apj, in press
\reference Terebey, S., Chandler, C. J., \& Andr\'e, P. 1993, \apj, 414,
759
\reference van der Tak, F. F. S. 2002, these proceedings
\reference van der Tak, F. F. S., van Dishoeck, E. F., Evans, N. J., II,
\&
   Blake, G. A. 2000, \apj, 537, 283
\reference Williams, J. P., \& McKee, C. F. 1997, \apj, 476, 166
\reference Young, C. H., Shirley, Y. L., Evans, N. J., II, \& Rawlings, 
    J. M. C. 2001, in prep.
\reference Zucconi, A., Walmsley, C. M., \& Galli, D. 2001, \aap, 376, 650


\end{references}
\end{document}